\documentclass[journal,twoside,web]{ieeecolor}
\usepackage{lcsys}
\usepackage{cite}
\usepackage{amsmath,amssymb,amsfonts,amstext}
\usepackage{graphicx}
\usepackage{textcomp}
  \def\BibTeX{{\rm B\kern-.05em{\sc i\kern-.025em b}\kern-.08em
     T\kern-.1667em\lower.7ex\hbox{E}\kern-.125emX}}
\usepackage[utf8]{inputenc}
\usepackage{balance}
\usepackage{comment}

\usepackage{soul}
\usepackage{subcaption}

\usepackage{mathtools}
\usepackage{float}

\usepackage{amsthm}

\newtheorem{definition}{Definition}

\newtheorem{remark}{Remark}

\newtheorem{problem}{Problem}

\usepackage{algorithm}
\usepackage{algorithmicx} 
\usepackage{algpseudocode} %

\DeclareMathOperator*{\minimize}{Minimize}
\DeclareMathOperator*{\maximize}{Maximize}

\makeatletter
\let\NAT@parse\undefined
\makeatother
\usepackage{hyperref}
\hypersetup{
  colorlinks=true,
    linkcolor= blue,
    allcolors=blue,
    citecolor = blue,
    filecolor=black,      
    urlcolor=blue,
    }

\title{Routing in Mixed Transportation Systems for Mobility Equity}

\author{Heeseung Bang$^{1,2}$, \textit{Student Member, IEEE}, Aditya Dave$^{2}$, \textit{Member, IEEE},\\Andreas A. Malikopoulos$^{2}$, \textit{Senior Member, IEEE}
    \thanks{This research was supported by NSF under Grants CNS-2149520 and CMMI-2219761.}
    \thanks{$^{1}$Department of Mechanical Engineering, University of Delaware, Newark, DE 19716, USA.} 
    \thanks{$^{2}$School of Civil and Environmental Engineering, Cornell University, Ithaca, NY 14850, USA. {\tt\small Email: \{hb489,a.dave,amaliko\}@cornell.edu}}
}

\date{September 2023}

\begin{document}

\maketitle
\thispagestyle{empty}  
\begin{abstract}

This letter proposes a routing framework in mixed transportation systems for improving mobility equity.
We present a strategic routing game that governs interactions between compliant and noncompliant vehicles, where noncompliant vehicles are modeled with cognitive hierarchy theory. Then, we introduce a mobility equity metric (MEM) to quantify the accessibility and fairness in the transportation network. We integrate the MEM into the routing framework to optimize it with adjustable weights for different transportation modes. The proposed approach bridges the gap between technological advancements and societal goals in mixed transportation systems to enhance efficiency and equity. We provide numerical examples and analysis of the results.
\end{abstract}


\section{Introduction}

\PARstart{D}{ue} to ongoing global urbanization and burgeoning urban populations, our society now faces not only the challenges of traffic congestion but also the associated societal issues, such as disparities in transportation opportunities, reduced accessibility to essential services for marginalized communities, and increased social isolation due to lengthy commutes.
Emerging mobility systems have received significant attention as a solution that can mitigate congestion, enhance safety, improve comfort, and optimize costs.

Numerous studies have addressed the coordination of connected and automated vehicles (CAVs) to achieve efficient operational methods in emerging mobility systems.
For example, a series of research papers addressed coordination problems in different traffic scenarios such as lane-changing \cite{wang2019q}, merging on-ramps in mixed traffic \cite{Nishanth2023AISmerging,xiao2021decentralized}, signalized intersection \cite{Malikopoulos2020}, and roundabouts \cite{chalaki2020experimental}.
These results have also been extended to the network level with vehicle-flow optimization.
Research efforts have also addressed various congestion-aware routing strategies considering mixed traffic contexts \cite{wollenstein2021routing} or targeting electric vehicles \cite{bang2021AEMoD}. Some approaches combine efficient routing with coordination strategies \cite{Bang2023flowbased,Bang2022combined} or learned traveler preference to achieve social objectives \cite{biyik2019green}.
However, exploring effective operational strategies that can mitigate the societal challenges of emerging mobility systems remains an open question.

The primary component of the societal challenges in emerging mobility systems is the uneven distribution of modes of transportation and accessibility to various urban resources. In response, research initiatives have arisen to address concerns of \textit{mobility equity} in a transportation network. As a concept, mobility equity has been examined from many diverse perspectives.
Notably, research has delved into socioeconomic parity across different strata of society, spatially equitable allocation of infrastructure, and the distribution of resources aligned with societal needs (for a detailed overview, see \cite{guo2020systematic}).
For instance, some studies examined the impact of individual characteristics, i.e., personal needs and abilities, on the effectiveness of the equity analysis \cite{dixit2020capturing}.
Meanwhile, other studies explored the link between social exclusion and transport disadvantages in accessibility \cite{schwanen2015rethinking} or provided a way of examining equity based on the transport choices \cite{chan2023choice}.
These investigations underscore the urgency of creating transportation systems that cater to the needs of all segments of society, enhancing accessibility and social inclusivity.
Despite the discrete examination of these challenges, however, there still needs to be more comprehensive efforts to interlink mobility equity with emerging mobility systems. Integrating the principles of mobility equity into the realm of emerging transportation modes still presents a gap in the existing body of knowledge.

To resolve these challenges, in this letter, we propose a routing framework in mixed transportation systems, where human-driven vehicles co-exist with CAVs, to improve mobility equity.
Our approach addresses different modes of transportation, accommodating private vehicles with varying levels of compliance.
First, we formulate a routing framework that suggests system-optimal solutions tailored explicitly to compliant vehicles.
To account for noncompliant vehicles' movement, we leverage the cognitive hierarchy model \cite{camerer2004cognitive}, inspired by many studies that have applied the cognitive hierarchy model to predict human decisions in transportation systems.
For example, Li et al. \cite{li2018game} utilized the cognitive hierarchy model in a game theoretic approach to manage the interaction between automated vehicles and human-driven vehicles. 
More recently, Feng and Wang \cite{feng2023strategic} utilized a cognitive hierarchy model to predict acceptance or rejection of the drivers in on-demand platforms.

By incorporating a cognitive hierarchy model, we can design a strategic game that governs interactions between compliant and noncompliant vehicles within the transportation system. Moreover, we introduce a mobility equity metric (MEM) to provide a quantifiable assessment of mobility equity. This metric captures essential aspects of accessibility, accounting for both geographical distances and monetary costs.
We then derive the MEM optimization problem integrated with the routing framework.
We solve this problem by adjusting the weights assigned to different transportation modes.
This comprehensive framework not only accounts for compliance variations and individual travel preferences but also integrates the overarching goal of equity into the decision-making process.

The main contributions of this work are the
(1) introduction of the MEM that can help resolve difficulties resulting from a lack of standard/clear metrics for mobility equity \cite{cantilina2021approaches};
(2) formulation and solution of the MEM optimization integrated with a multi-modal routing framework, which offers a unique perspective on mobility optimization and equity enhancement and
(3) development of variations of the MEM that support the analysis in small networks.

The remainder of this paper is organized as follows.
In Section \ref{sec:routing}, we formulate a routing problem within mixed transportation systems.
We introduce a mathematical definition of MEM in Section \ref{sec:mem} and present the MEM optimization framework in Section \ref{sec:memOpt}.
Section \ref{sec:simulation} showcases a numerical example with practical implementation strategies on a smaller network and subsequently analyzes results.
Finally, in Section \ref{sec:conclusion}, we provide concluding remarks and discuss directions for future research.


\section{Routing in Mixed Transportation} \label{sec:routing}

This section presents a routing framework of a mixed transportation system with various modes of transportation, including public transportation and privately owned vehicles. The private vehicles are comprised of human-driven vehicles and CAVs.
We assume that public transportation and CAVs follow the system's route suggestion while private human-driven vehicles may or may not comply with the suggestion.
We seek to provide system-wide optimal suggestions to all compliant vehicles, including public transportation, CAVs, and private human-driven vehicles. However, such suggestions must account for non-compliant private vehicles (NPVs) decisions. Thus, we distinguish between compliant private vehicles (CPVs) and NPVs in our formulation.

Consider a road network given by a directed graph $\mathcal{G} = (\mathcal{V},\mathcal{E})$, where $\mathcal{V}\subset\mathbb{N}$ is a set of nodes and $\mathcal{E}\subset\mathcal{V}\times\mathcal{V}$ is a set of edges.
We consider $\mathcal{O}$ and $\mathcal{D}$ as the set of origins and destinations, respectively.
Let $\mathcal{N}=\{1,\dots,N\}$, $N\in\mathbb{N},$ denote a set of trips, each comprising of an origin-destination pair, and $\mathcal{M}$ a set of modes of transportation available for system-wide routing, e.g., public transportation, shared mobility, private vehicles.
For each trip $n \in \mathcal{N}$, information of origin $o_n\in\mathcal{O}$, destination $d_n\in\mathcal{D}$, and compliant travel demand rate $\alpha_{m,n}\in\mathbb{R}_{>0}$ for each mode $m\in\mathcal{M}$ is given.
We let $x^{ij}_{m,n} \in \mathbb{R}_{\geq0}$ be the flow on edge $(i,j) \in \mathcal{E}$ traveling for the trip $n$ with the specific mode $m$. Then, the total complying-vehicle flow is given by $x^{ij}=\sum_m\sum_n x^{ij}_{m,n}$.
Note that we distinguish the flow of NPVs on edge $(i,j)$ by denoting it with $q^{ij} \in \mathbb{R}_{\geq0}$.
Given both the total complying and noncomplying flows on edge $(i,j)$, we estimate travel time using the \textit{Bureau of Public Roads (BPR)} function denoted by
\begin{equation}
    t^{ij}(x^{ij}+q^{ij}) = t^{ij}_0 \cdot \left( 1+0.15 \left(\frac{x^{ij}+q^{ij}}{\gamma^{ij}}\right)^4\right),
\end{equation}
where $t^{ij}_0$ is the free-flow travel time and $\gamma^{ij}$ is capacity of the road on edge $(i,j)$.

Recall that our system can provide socially equitable suggestions only to compliant vehicles willing to accept those while NPVs seek to maximize their utility strategically.
The routing process resulting from the interactions between these entities can be formulated as a game.
Next, we describe the system-centric optimization problem for compliant vehicles and follow it with the decision-making model for NPVs to describe the interactions within this game.

\textit{1) System-Centric Routing:} To suggest socially-equitable routes to the complying-vehicle flow, we solve the following flow optimization problem.
\begin{problem}[System-Centric Routing] \label{pb:system-centric}
    \begin{equation}
        \begin{aligned}
            \minimize_{\{x^{ij}_{m,n}\}} ~& \sum_{m\in\mathcal{M}} w_m \left\{ \sum_{n\in\mathcal{N}}\sum_{(i,j)\in\mathcal{E}} t^{ij}(x^{ij}+q^{ij})\cdot x^{ij}_{m,n} \right\}\\
            \mathrm{subject~to:~} & \sum_{k:(j,k)\in\mathcal{E}} x^{jk}_{m,n} = \alpha_{m,n},\\
            &\hspace{17ex}\forall m\in\mathcal{M},n\in\mathcal{N},j=o_n,\\
            & \sum_{i:(i,j)\in\mathcal{E}} x^{ij}_{m,n} = \alpha_{m,n},\\
            &\hspace{17ex}\forall m\in\mathcal{M},n\in\mathcal{N},j=d_n,\\
            & \sum_{i:(i,j)\in\mathcal{E}} x^{ij}_{m,n} = \sum_{k:(j,k)\in\mathcal{E}} x^{jk}_{m,n},\\
            &\hspace{10ex}\forall m\in\mathcal{M},n\in\mathcal{N},j\in\mathcal{V}\setminus\{o_n,d_n\},            
        \end{aligned}
    \end{equation}
    where $w_m$ is the weight for transportation mode $m$.
\end{problem}
The constraints ensure the flow matches the demand rate and connects corresponding origins and destinations.
Problem \ref{pb:system-centric} is a convex problem as the BPR function is convex in its domain and constraints are linear.

\begin{figure}
    \centering
    \includegraphics[width=0.45\linewidth]{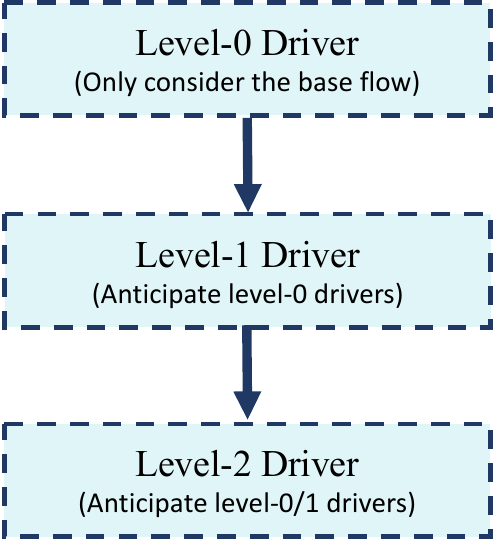}
    \caption{Conceptual diagram of cognitive hierarchy model.}
    \label{fig:hierarchyModel}
\end{figure}

\textit{2) Strategic Rouing for NPVs:} In recent research articles, NPVs have been modeled as one group whose interactions result in a Wardrop equilibrium (see \cite{beckmann1956studies} and references therein).
Wardrop equilibrium is reached at steady state after transient phases wherein travelers continuously adjust their route selection.
In a real traffic scenario, however, it requires perfectly rational drivers with access to all information of other NPVs or a significant amount of time for drivers to interact with each other and reach equilibrium.
Therefore, as an alternative, we introduce the cognitive hierarchy model (see Fig. \ref{fig:hierarchyModel}) to describe the behavior of NPVs.
This model categorizes human drivers into different levels of decision-making rationality.
At each level, human drivers can anticipate lower-level drivers' decisions and make ``smarter" decisions.
For instance, level-0 drivers decide based on publicly available information. Meanwhile, level-1 drivers anticipate level-0 drivers' decisions and select better paths, and level-2 drivers can anticipate level-0 and level-1 drivers.
According to the experimental results, humans can most commonly anticipate others' decisions up to level-2 \cite{costa2009comparing,costa2006cognition}; thus, we restrict our model to level-2 decision-making.

\begin{remark}
    The cognitive hierarchy model is suitable for considering the behavior of NPVs when the percentage of compliant vehicles is high. This can occur in the presence of a large number of CAVs and public transportation. In this case, the traffic flow generated by compliant vehicles can represent the traffic situation across the network, thus making it reasonable to assume that a level-0 driver cannot anticipate behaviors beyond this information. 
    However, if the majority of the drivers are noncompliant, this would result in a massive gap between their anticipation and experience of traffic on the roads. In this situation, they can potentially learn the dynamics and try to anticipate others' decisions.
\end{remark}

For each trip $n \in \mathcal{N}$, we also have the demand rate for each $\ell$-level NPV, $\ell=0,1,2,$ from the origin $o_n \in \mathcal{O}$ to the corresponding destination $d_n \in \mathcal{D}$.
For an $\ell$-level NPV traveling for trip $n$, we define assignment vector $A_{\ell,n}\in 2^{|\mathcal{E}|}$ where the element $a^{ij}_{\ell,n}$ takes value of $1$ if the $\ell$-level NPV for trip $n$ uses the edge $(i,j)$ and takes value of $0$ otherwise. We solve the following problem for each NPV to determine their path.

\begin{problem}[Strategic Routing of $\ell$-Level NPV] \label{pb:selfish}
\begin{equation}
        \begin{aligned}
        \minimize_{\{a_{\ell,n}^{ij}\}}~& \sum_{n\in\mathcal{N}}\sum_{(i,j)\in\mathcal{E}} t^{ij}\left(x^{ij}+\sum_{l=0}^{\ell-1}q_{l}^{ij}\right)\cdot a_{\ell,n}^{ij} \\
        \mathrm{subject~to:~} & \sum_{k:(j,k)\in\mathcal{E}} a^{jk}_{\ell,n} = 1,~\forall n\in\mathcal{N},j=o_n,\\
        & \sum_{i:(i,j)\in\mathcal{E}} a^{ij}_{\ell,n} = 1,~\forall n\in\mathcal{N},j=d_n,\\
        & \sum_{i:(i,j)\in\mathcal{E}} a^{ij}_{\ell,n} = \sum_{j:(j,k)\in\mathcal{E}} a^{jk}_{\ell,n},\\
        &\hspace{13ex}\forall n\in\mathcal{N},j\in\mathcal{V}\setminus\{o_n,d_n\},
        \end{aligned}
    \end{equation}
    where $q_\ell^{ij} = \sum_{n\in\mathcal{N}}q_{\ell,n} \cdot a_{\ell,n}^{ij}$.
\end{problem}

Problem \ref{pb:selfish} is a routing problem for a single NPV, which can be solved using any graph search algorithm such as Dijkstra or A$^*$.
As each $\ell$-level NPV only anticipates lower-level NPVs, all NPVs at the same level make the same routing decision.

Note that $q^{ij}$ in Problem \ref{pb:system-centric} is the flow of NPVs, and $x^{ij}$ in Problems \ref{pb:system-centric} and \ref{pb:selfish} is the flow of all the compliant vehicles.
As the problems are coupled and affect each other, they form a strategic game.
To resolve this game, we iteratively solve them until the solutions converge to an equilibrium.
Since there is no theoretical guarantee of convergence, we provide a way to induce convergence in Section \ref{subsec:convergence}.

\begin{remark} \label{remark_1}
    The solution to Problem \ref{pb:system-centric} is sensitive to the choice of weights.
    Therefore, it is required to have principled ways of selecting appropriate weights. In the next section, we introduce a mobility equity metric that evaluates the accessibility and equity of transportation resources in a network. This metric will allow us to select socially appropriate weights.
\end{remark}


\section{Mobility Equity Metric} \label{sec:mem}

Mobility equity refers to the fair distribution of transportation resources and opportunities among diverse communities, regardless of socioeconomic status, location, or other factors.
Although people are bringing attention to mobility equity, there has yet to be a strict definition and a convention on its effective quantification.
In this section, we propose a mobility equity metric (MEM) that quantifies the degree to which transportation services cater to the needs of different demographic groups, highlighting any discrepancies in access. In this metric, we aim to account for various factors pertaining to mobility, such as accessibility to essential services (e.g., healthcare, education, and employment), affordability, travel time, and availability of multiple modes of transportation. Next, we present the mathematical definition.

Let $\mathcal{M}$ be the set of all modes of transportation, $\mathcal{S}$ be the set of essential services, and $\kappa$ be the price sensitivity.
For each $m\in\mathcal{M}$ and $s\in\mathcal{S}$, we let $c_m$ denote the cost per passenger mile of utilizing transportation mode $m$, $\beta^s$ denote the priority level of service $s$, and $\sigma_m^s(\tau_m)$ denote the average number of services accessible within time threshold $\tau_m$ from all selected origins in the network.

\begin{definition} \label{def:mem}
For a given transportation network $\mathcal{G}$ with modes in $\mathcal{M}$ and services in $\mathcal{S}$, the mobility equity metric is
\begin{equation} \label{eq:MEM}
    \textit{MEM} = \sum_{m\in\mathcal{M}} e^{-\kappa c_m} \cdot \left\{ \sum_{s\in\mathcal{S}} \beta^s \sigma_m^s(\tau_m) \right\}.
\end{equation}
\end{definition}

Here, $e^{-\kappa c_m}$ ensures that MEM decreases with an increase in the cost per passenger mile, and $\beta^s \sigma_m^s(\tau_m)$ ensures that MEM increases with respect to an increase of accessibility to the essential services. These terms collectively prioritize increasing access to services at lower costs to passengers to increase MEM.

\begin{remark}
    An advantage of the MEM defined in \eqref{eq:MEM} is that, in practice, it can be computed purely using publicly available data.
    For example,  the base flows in a traffic network can be measured over time, the number of services $\sigma_m^s(\tau_m)$ can be counted using an isochrone map for the base traffic conditions, and the costs of transportation can be computed from travel times and fuel consumption.
\end{remark}

\begin{remark}
    Recall that $\sigma_m^s(\cdot)$ represents the average number of accessible services from selected origins in the network.
    We anticipate that by selecting these origins carefully to include diverse social groups, it is possible to consider the impact of social factors on mobility equity. Subsequently, the MEM can facilitate a fair distribution of transportation resources.
\end{remark}

Next, we formulate an optimization problem to integrate the MEM from Definition \ref{def:mem} into the routing framework presented in Section \ref{sec:routing}.
This allows us to develop a mobility-equity-focused approach to select the weights to prioritize various modes of transportation in system-centric routing, as described in Remark \ref{remark_1}.

\section{Mobility Equity Optimization} \label{sec:memOpt}

In this section, we formulate the MEM optimization problem in the mixed-transportation network.
The problem aims to maximize the MEM by improving accessibility to the services with cost-efficient modes of transportation.
Here, accessibility is captured by counting the number of accessible services within a time threshold.
In our routing formulation, solutions to Problem \ref{pb:system-centric} and Problem \ref{pb:selfish} will be the net flow on the network, which can be used to estimate travel time for given origins and destinations.
Note that the net flow is determined for given weights $w$.
Thus, we can formulate the MEM optimization problem with respect to the weights.

\begin{problem}[Mobility Equity Maximization] \label{pb:mem}
\begin{equation}
    \begin{aligned}
        \maximize_{w}&~~ \textit{MEM}\\
        \mathrm{subject~to:}&~~ \delta^\mathrm{pv}(w) \leq \gamma,
    \end{aligned}
\end{equation}
where $\delta^\textrm{pv}$ is the average travel-time difference between CPVs and NPVs, and $\gamma$ is the upper limit of the difference.
\end{problem}

\begin{figure}
    \centering
    \includegraphics[width=\linewidth]{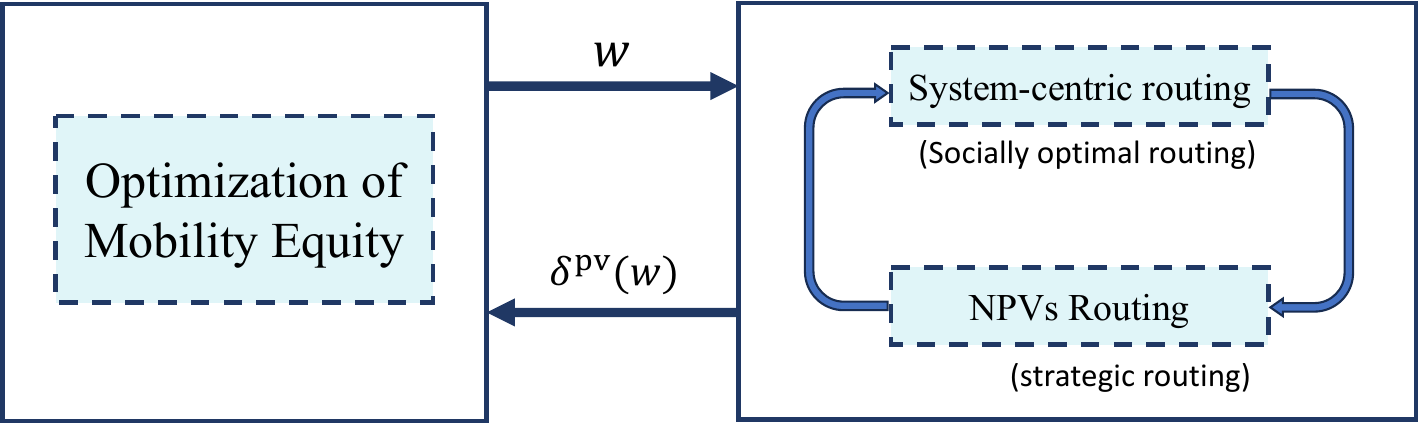}
    \caption{Structure of the socially-optimal routing problem.}
    \label{fig:routing_prob}
\end{figure}

Figure \ref{fig:routing_prob} illustrates the structure of integrating routing and MEM optimization.
For each possible $w$, it is required to solve Problem \ref{pb:system-centric} and Problem \ref{pb:selfish} repeatedly until their solution converges. 
Therefore, Problem \ref{pb:mem} can only be solved numerically.
We impose a constraint on the travel-time difference $\delta^\textrm{pv}$ between CPVs and NPVs because the routing framework would sacrifice CPVs' travel time to maximize MEM.
Without this constraint, CPVs would no longer comply if their time loss increases to a certain level.
Thus, we impose the upper bound $\gamma$ in order to keep their time loss bearable.

\begin{remark}
    In practice, CPVs in the system-centric routing tend to have longer travel times than the NPVs.
    Thus, their disincentive towards complying is captured by the differences in travel time between CPVs and NPVs, which we bound by the constant $\gamma$. Though we do not explicitly explore this direction, the constant $\gamma$ should be determined through monetary incentives provided to drivers to maintain compliance.
\end{remark}


\subsection{Inducing Convergence} \label{subsec:convergence}

One potential cause of challenge to solve Problem \ref{pb:mem} is a lack of convergence of flows in the routing game (right-hand side of Fig. \ref{fig:routing_prob}) induced by a specific choice of weights.
In this subsection, we explain the ``chattering behavior" in the routing game and propose a resolution to this concern by controlling the flows of compliant vehicles in the system-centric problem. 
The chattering behavior may originate from the fact that to optimize the MEM, the system-centric routing problem may receive weights prioritizing the travel time for public vehicles with a smaller cost per passenger mile over CPVs.
To understand this phenomenon, consider a single origin and destination with two possible paths $p_1$ and $p_2$. For the compliant vehicles, the optimal solution to Problem \ref{pb:system-centric} is to assign a shorter-time path $p_1$ to the public transportation and a longer-time path $p_2$ to the CPVs. 
Then, NPVs select $p_1$ for their benefit because $p_1$ is still the shortest-time path.
This results in a scenario where public transportation and NPVs travel in traffic congestion while CPVs travel using a longer but less crowded path.
In the next iteration, compliant vehicles would thus be assigned different paths so that public transportation can travel faster.
In response to these flow changes, NPVs would also change their decision to travel on the same path as public transportation because it would always be the less congested path.
Due to the repeated nature of these interactions, routing decisions may chatter over time as the system-centric routing attempts to prioritize public vehicles and NPVs keep following.

In case chattering occurs, we impose an addition constraint given by
    $\sum_m \sum_n x^{ij}_{m,n} = f^{ij}$,
where $f^{ij}$ is the total compliant vehicles flow on edge $(i,j)$ at the previous iteration.
This constraint ensures compliant vehicle flow is the same as the previous iteration while improving public transportation travel time.
Although compliant vehicles changed their paths, there is no incentive for NPVs to change their path because the total flow on the roads remains the same.

\section{Numerical Implementation} \label{sec:simulation}

\begin{figure}
    \centering
    \includegraphics[width=0.85\linewidth]{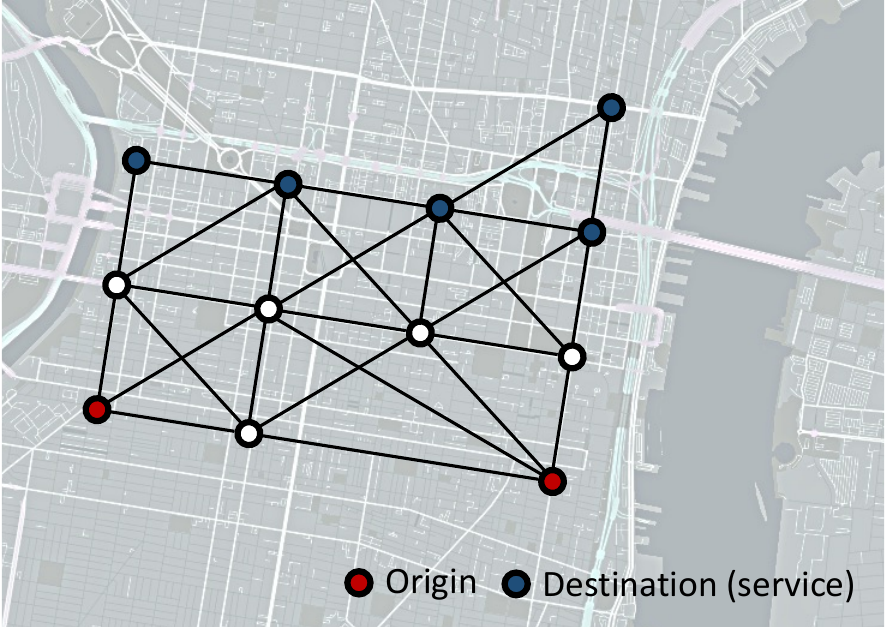}
    \caption{Small network for simulation with two origins and five destinations.}
    \label{fig:toyNetwork}
\end{figure}

In this section, we provide a numerical implementation and analysis of the results.
To prove our concept, we consider a small network with $12$ nodes and $54$ edges, as illustrated in Fig. \ref{fig:toyNetwork}.
We introduce $10$ travel demands ($2$ origins and $5$ destinations) and randomly generated demand rates, where origins and destinations are considered as residential areas and essential services, respectively.
In our implementation, we address a single type of service, i.e., $|\mathcal{S}|=1$, and mode of transportation given by $\mathcal{M}=\{\mathrm{public~transportation},\mathrm{CPVs}\}$.
We assume that travel demands exist for all possible origin-destination pairs and that the demand rates are known a priori.

To evaluate MEM, we need the average number of services accessible within the time threshold $\tau_m$ given by
\begin{equation} \label{eqn:sigma}
    \sigma_m^s(\tau_m) := \frac{\sum_{o\in\mathcal{O}} \alpha_m^o \cdot \sum_{d\in\mathcal{D}}\mathbb{I}\Big[t_m^{o,d} \leq \tau_m\Big]}{\sum_{o\in\mathcal{O}} \alpha_m^o},
\end{equation}
where $\alpha_m^o$ is the compliant travel demand rate departing at $o$ using mode $m$, $\mathbb{I}[\cdot]$ is the indicator function yielding the number of services within $\{o,d\}$, and $t_m^{o,d}$ is an average travel time for $o$ to $d$ via mode $m$.
The number of services counted with $\sum_{d\in\mathcal{D}}\mathbb{I}[t_m^{o,d} \leq \tau_m]$ is weighted with the flow $\alpha_m^o$ to consider the different levels of influence for different travel demands.
Here, we determine $t_m^{o,d}$ by
\begin{equation}
t_m^{o,d}=\frac{\sum_{(i,j)\in\mathcal{E}}t^{ij}(x^{ij}+q^{ij})\cdot x^{ij}_{m,n}}{\sum_{(i,j)\in\mathcal{E}} x^{ij}_{m,n}},
\end{equation}
where $n\in\mathcal{N}$ is the travel corresponding to the origin-destination pair $(o,d)$.

\subsection{Continuous Approximation of Mobility Equity Metric}

For a network with numerous origins and destinations, using the indicator $\mathbb{I}$ produces meaningful changes to the MEM because small shifts in travel time result in a change in the number of services accessible within a time threshold.
In contrast, for small networks, using the indicator function $\mathbb{I}$ to capture the accessibility may barely affect MEM for variations in travel time. 
To produce meaningful results for a small network, we approximate the indicator function with a continuous function that is more sensitive to changes in travel time.
The approximate function is given by
\begin{equation}
    \Tilde{\mathbb{I}}(t) = 1-\frac{1}{1+e^{-k(t-\tau_m)}}, \label{eqn:approx_util}
\end{equation}
where $k\in\mathbb{R}_{>0}$ is a parameter of the slope. 
As $k$ increases, the function gets closer to the original indicator function, while it becomes sensitive to the input as $k$ decreases.
By adopting \eqref{eqn:approx_util} instead of indicator function $\mathbb{I}$, \eqref{eqn:sigma} can provide distinguishable outcomes even in a small network.

\subsection{Numerical Simulation Results}

\begin{figure*}[t!]
    \centering
    \begin{subfigure}{0.325\linewidth}
        \centering
        \includegraphics[width=\linewidth]{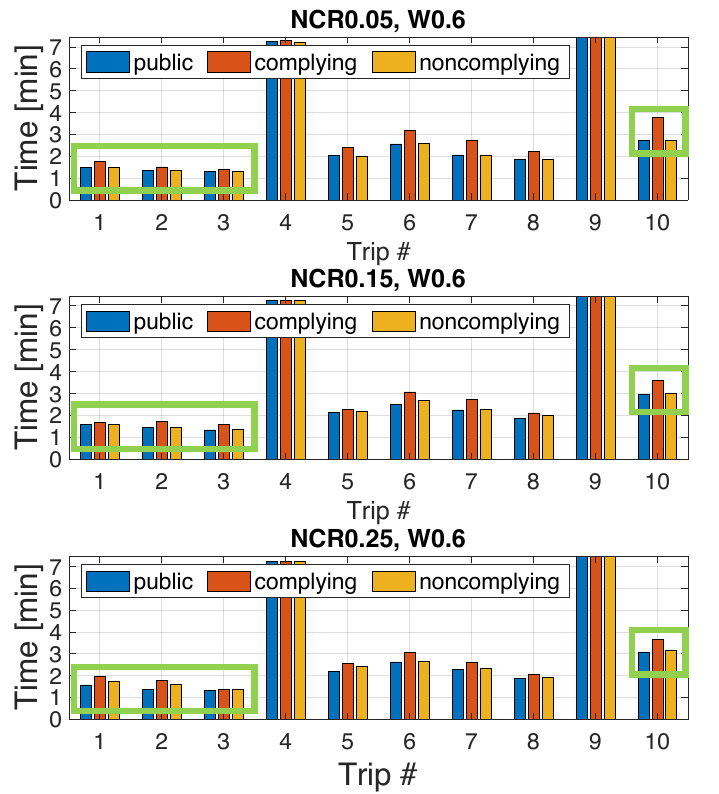}
        \caption{Different noncompliance rates}
        \label{fig:diff_n}
    \end{subfigure}%
    \begin{subfigure}{0.325\linewidth}
        \centering
        \includegraphics[width=\linewidth]{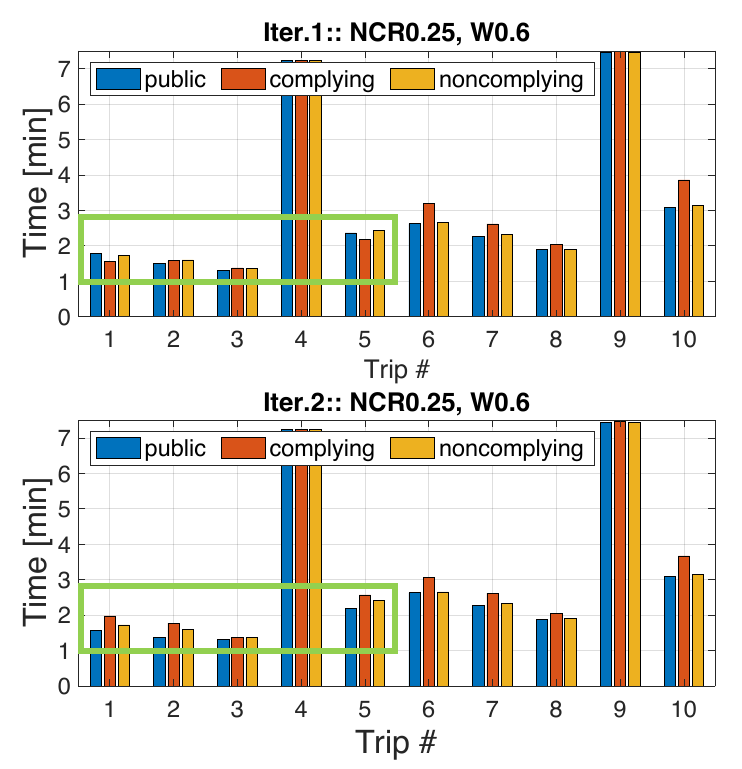}
        \caption{Different iterations}
        \label{fig:diff_i}
    \end{subfigure}
    \begin{subfigure}{0.325\linewidth}
        \centering
        \includegraphics[width=\linewidth]{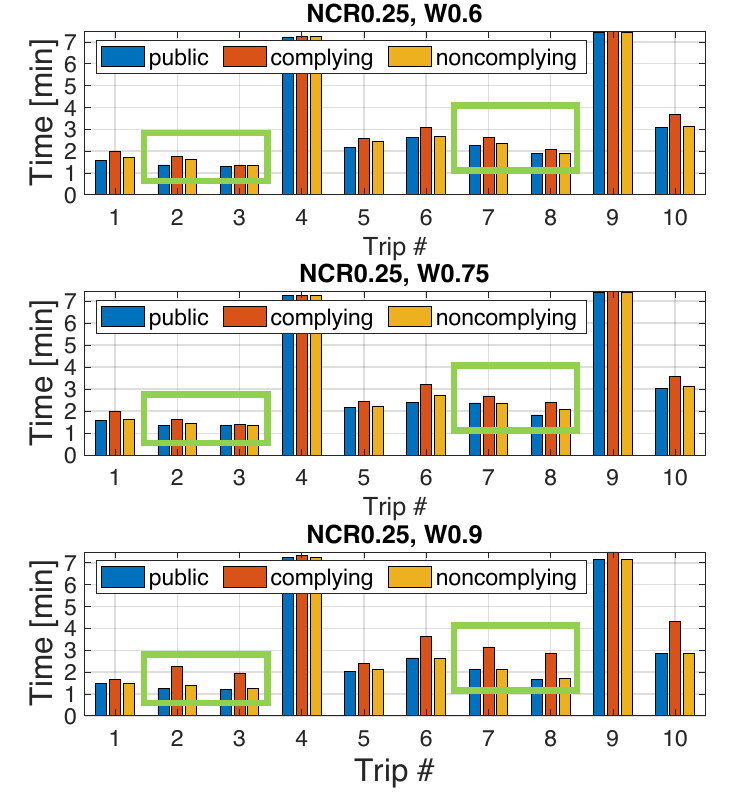}
        \caption{Different weights}
        \label{fig:diff_w}
    \end{subfigure}
    \caption{Travel time of each demand and transportation mode for different variables.}
    \label{fig:sim_travelTime}
\end{figure*}

Given the network and travel demands, we first ran simulations to analyze the effect of the compliance rate on the MEM.
Figure \ref{fig:diff_n} illustrates the travel time of each transportation mode for different trips for different noncompliance rates (NCR).
The figure shows that the overall travel time increases with the NCR because increasing NPVs generates traffic congestion and reduces the benefit of system-centric routing.

Figure \ref{fig:diff_i} shows the simulation for the first and second routing iterations.
For public transportation with longer travel time than CPVs, our routing framework modified the flow so that CPVs yield the roads to public transportation without affecting NPVs' travel time.
This result shows that our method successfully modified the flow without incentivizing NPVs to deviate from their previous routing decisions.

\begin{figure}
    \centering
    \includegraphics[width=0.8\linewidth]{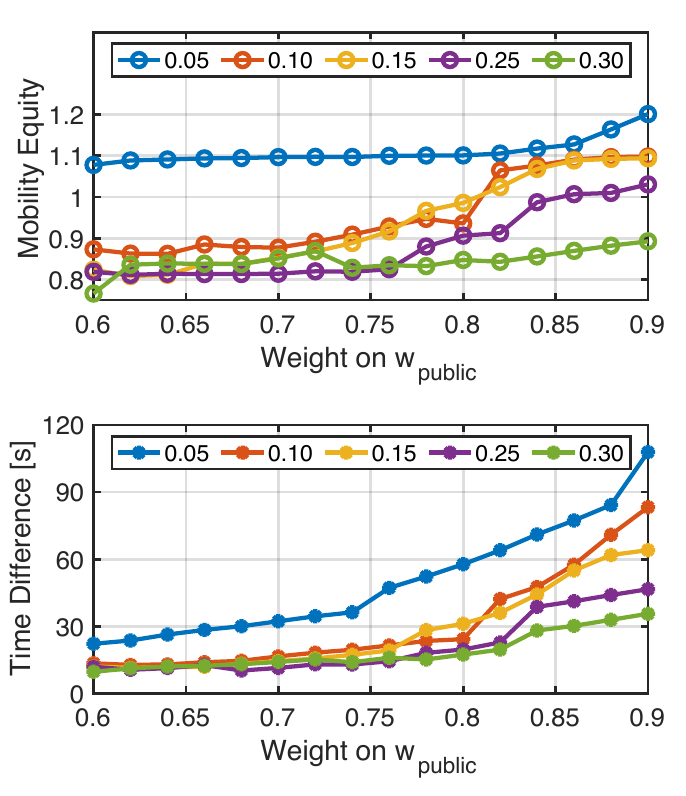}
    \caption{Mobility equity metric and time difference for different noncompliance rate.}
    \label{fig:sim_mem}
\end{figure}

Next, we conducted simulations for different weights for the modes of transportation.
Figure \ref{fig:diff_w} is the simulation results at a microscopic level for three different weights.
It shows that the travel-time difference between CPVs and NPVs increases as the weight of public transportation increases.
This aligns with the intuition wherein the CPVs increasingly sacrifice their travel time as the system prioritizes public transportation.
Figure \ref{fig:sim_mem} illustrates MEM and the time difference $\delta^\textrm{pv}(w)$ for both different weights and different noncompliance rates.
Overall, as the weight on public transportation increased, the MEM and the time difference have increased.
This tendency has appeared because both public transportation and NPVs benefited in travel time at the expense of CPVs' travel time.
Moreover, MEM increased as the noncompliance rate decreased because more vehicles were involved in system-centric routing.
For specific noncompliance rates, there exist points where MEM dramatically jumps while the time difference slightly increases.
Thus, one can account for the results and provide incentives using a mechanism design to increase the limit and enhance MEM even more.



\section{Concluding Remarks} \label{sec:conclusion}

In this letter, we presented a routing framework in a mixed transportation system for improving mobility equity in the network.
We formulated a routing game between all compliant vehicles and NPVs. Then, we proposed MEM and formulated the MEM optimization problem.
We presented a numerical example and implementation method that yields meaningful results in a small network.
Through the simulations, we verified that our framework works for improving MEM.
Future work should consider using MEM in real-road networks and account for the effect of compliance rate by designing monetary incentives.


\bibliographystyle{IEEEtran}
\bibliography{Bang,IDS}

\end{document}